\documentclass{aastex62}
\usepackage[utf8]{inputenc}
\usepackage{graphicx}
\usepackage{natbib}
\usepackage{float}
\usepackage{amssymb,amsmath}
\begin{document}

\title{A Fast Approximate Approach to Microlensing Survey Analysis}
\author{Somayeh Khakpash, Matthew Penny, Joshua Pepper}

\section{Abstract}

Microlensing can be used to discover exoplanets of a wide range of masses with orbits beyond $\sim 1$  AU, and even free-floating planets. The WFIRST mission will use microlensing to discover approximately 1600 planets by monitoring  $\sim$ 100 million stars to find  $\sim$ 50000 microlensing events. Modelling each microlensing event, especially the ones involving two or more lenses, is typically complicated and time-consuming, and analyzing thousands of WFIRST microlensing events is possibly infeasible using current methods. Here, we present an algorithm that is able to rapidly evaluate thousands of simulated WFIRST binary-lens microlensing light curves, returning an estimate for the physical parameters of the lens systems. We find that this algorithm can recover projected separations between the planet and the star very well for low-mass-ratio events, and can also estimate mass ratios within an order of magnitude for events with wide and close caustic topologies.

\section{Introduction}
Today about 3700\footnote{https://exoplanetarchive.ipac.caltech.edu} exoplanets have been discovered with a variety of different methods \citep{akeson2013nasa}. Among all methods, the microlensing technique is the only one that has the sensitivity to detect low-mass planets in orbits beyond $\sim1$ AU \citep{2012ARA&A..50..411G}. The Wide Field Infrared Survey Telescope (WFIRST) is a NASA mission expected to be launched in the mid 2020's. It consists of a 2.4 m telescope and a wide field camera with a field of view that is 100 times greater than that of the Hubble Space Telescope. One of the main scientific goals of this mission is searching for exoplanets using microlensing observations in the Galactic bulge \citep{spergel2015wide}.

Analyzing the data from the WFIRST microlensing survey presents a significant computational challenge. Out of $\sim10^8$ stars to be observed in the WFIRST microlensing survey, there there is expected to be $\sim54,000$ microlensing events detected, among which will be $\sim1600$ planetray signals, along with multiple types of stellar variability \citep{2018arXiv180802490P}.  It will be necessary to search through the full set of light curves to identify microlensing behavior, and then each individual microlensing event must be carefully analyzed to allow the detection of planets.

The goal of light curve modeling in microlensing is to find the best-fitting values of the seven parameters of the a binary microlensing light curve. Three of these parameters are the event timescale $t_E$, time of event $t_0$, and impact parameter $u_0$, which are also shared with a single lens event. Three, the mass ratio $q$, projected separation $s$, and angle of the source trajectory relative to the binary axis $\alpha$, describe the binary lens. The final parameter is $\rho$, the ratio of the angular diameter of the source to the angular Einstein ring. There are additional parameters that describe the brightness of the source star and any unrelated blended light, but are unimportant for this discussion. Higher order effects such as parallax and orbital motion require further parameters and can significantly complicate the analysis. As currently performed, the process of finding a best-fit function that describes the features seen in the light curve is usually done by ${\chi}^2$ minimization using a downhill simplex algorithm or Markov Chain Monte Carlo (MCMC) methods \citep{2010MNRAS.408.2188B}. However, these methods become challenging for binary-lens events, since the parameter space for these events is very large and has more than one minimum (see introduction of \citet{penny2014speeding} for a review). Also, the light curve of a binary-lens event can exhibit complicated features that make modeling these events difficult and time-consuming. Computation of the magnification at each timestep requires the numerical solution of multiple fifth-order polynomials  \citep[e.g.][]{gould2008hexadecapole}, especially when finite source effects are significant. A more insidious challenge is that the physical parameters of the system often do not have a simple mapping to the features of the light curve, making it difficult to select useful initial guesses for the fitting. Presently, initial guesses are often found by brute force grid searches of a subset of parameters (usually $q$, $s$ and $\alpha$), or by a brute force search of light curve morphologies \citep[e.g][]{2015MNRAS.450.1565L}. Often, the fitting process requires a good deal of human supervision. 

 These difficulties, and the expected order of magnitude increase in data volume compared to ground-base datasets, motivate the search for improvements to the algorithms used for modelling microlensing events. For this purpose, we have reexamined some ideas that were presented in one of the first papers to suggest using microlensing to search for planets. \citet{gould1992discovering} first introduced the idea that planet-star mass ratio $q$ can be measured directly from the light curve by finding the ratio of planetary perturbation duration to the lens star's Einstein timescale (see Equation \ref{eqn:find_q}). Later, \citet{gaudi1997planet} expanded this idea and discussed the degeneracies that would arise when one finds $q$ and $s$ by measuring parameters directly from the light curve. 

In this paper, we build upon these ideas by developing an automatable algorithm to extract estimates of the principle event parameters using only analytic functions. We test the effectiveness of our algorithm by quantitatifying its ability to estimate the parameters of a set of simulated WFIRST microlensing events. It is important to note that such a method can be applied to any set of high cadence microlensing light curves. This approach will then help selecting individual microlensing events for intensive analysis for precise parameter determination. 

In this project, we focus on the approximate extraction of physical parameters from binary-lens light curves. We assume that such light curves have been correctly identified from all survey light curves, which itself will be a non-trivial task. We apply this algorithm to a sample of simulated WFIRST light curves, and compare the results of the procedure to the true physical parameters of each system. In section \ref{sec:methods}, we introduce the model and the initial idea leading to the selection of this model and discuss the nature of our algorithm, then, we describe the simulated data in section \ref{sec:sim_data}. In section \ref{sec:results}, we show examples of the resulting parameter fits, and in section \ref{sec:conclusion} we discuss the results and present the applications of the algorithm and next steps of this project.

\section{Methods}\label{sec:methods}

\subsection{Gaudi \& Gould (1997) By-eye fitting}\label{subsec:by-eye fitting}

\citet{gaudi1997planet} expands on the idea of finding mass ratio from planetary duration \citep{gould1992discovering} and discusses how well the two planetary parameters $s$ and $q$ can be measured from features in the light curve. From a microlensing light curve, one can measure maximum magnification, time of maximum magnification and the event duration and from these quantities the three parameters of a single-lens event, $t_0$, $u_0$ and $t_E$ can be determined. Then, looking at the planetary anomaly, one can measure the time of the anomaly($t_p$), the duration of the anomaly ($t_{E,p}$), and its fractional deviation from the single-lens magnification ($a$ or amplitude). Finally, the planetary parameters $s$, $q$, and $\alpha$ can be found from the planetary anomaly features in the light curve. For the general purpose of this method, the blending of the source light with its neighbor stars is ignored \citep{gaudi1997planet,2012ARA&A..50..411G}.

This method relies on using the approximate duration of the planetary perturbation to estimate the planet mass ratio, and the time of the perturbation relative to the main event peak to estimate the projected separation, Therefore, the results are subject to a two-fold discrete degeneracy in $s$ and a continuous degeneracy in $q$. The discrete degeneracy in $s$ is caused when it is not clear if the planet is perturbing the major or minor image, and is usually easily solved when the light curve is fully covered. The continuous degeneracy in $q$ happens due to the presence of the finite source effects. The parameter $q$ is determined from the ratio of the planetary perturbation duration to the Einstein timescale, so that when the source size is larger than the planet Einstein ring, the planetary perturbation duration is determined by the source crossing time instead of the planet Einstein ring \citep{gaudi1997planet}.  

Additionally, since the parameter $s$ is calculated using the approximate time of the anomaly relative to the peak of the single-lens event, the result would only be acceptable for planet perturbations caused by planetary caustics and not the central caustic events. In the following subsections, we describe an approximate analytic light curve model that encodes these principles, and an algorithm to fit a light curve with this model and extract its parameters, and we also explain how the degeneracies in $s$ and $q$ challenge our calculations and how we have handled it.

\subsection{The model}\label{subsec:model}

\begin{figure}[H]
    \centerline{\includegraphics[width=12cm]{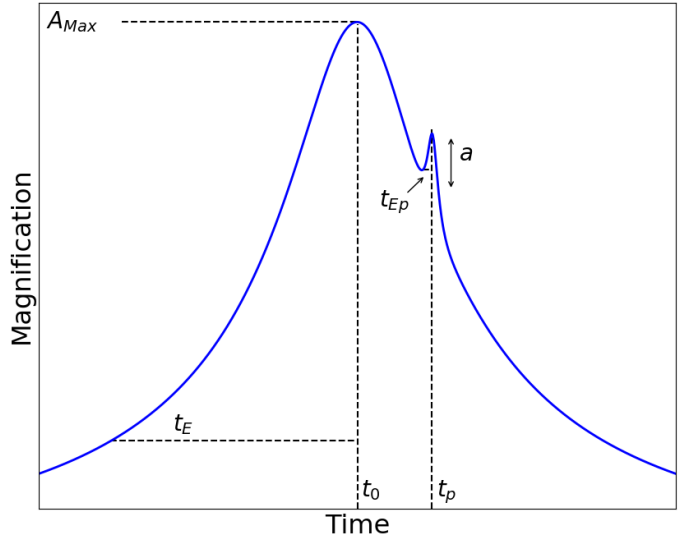}}
    \caption{Parameters for the heuristic model}
    \label{fig:inputparam}
\end{figure}

In this model, we assume that a planetary microlensing event can be approximated as a standard single lens model \citep{paczynski1986gravitational} with a smooth, short-lived Gaussian anomaly. This function is shown in Equations \ref{eqn:fs-1}, \ref{eqn:A-2} and \ref{eqn:A-3}, and Figure \ref{fig:inputparam} shows a schematic representation of how the parameters relate to the features in the light curve. 

\begin{equation}\label{eqn:fs-1}
    F(t) = f_s \times A(t)+(1-f_s) 
\end{equation}
\begin{equation}\label{eqn:A-2}
    A(t) = \frac{u(t)^2 + 2}{u(t)\sqrt{u^2 + 4} }+ a\:\exp\left[-\frac{{(t-t_p)}^2}{2t_{E,p}^2}\right]
\end{equation}
\begin{equation}\label{eqn:A-3}
    u(t) = \sqrt{{u_0}^2 + {\left(\frac{t-t_0}{t_E}\right)}^2}
\end{equation}

The Paczynski model includes four parameters: the time of the maximum magnification ($t_0$), the impact parameter ($u_0$), the Einstein crossing time ($t_E$), and the blending coefficient of the source ($f_s$).The Gaussian anomaly parameters are the mean, standard deviation, and the amplitude which we relate to the time of the planetary event ($t_p$), the Einstein timescale of the planetary event ($t_{E,p}$), and the height of the planetary event (Amplitude, or $a$), respectively (Figure \ref{fig:inputparam}). The quantity $f_s$ in Equation \ref{eqn:fs-1} is called the blending parameter, and is the ratio of the unmagnified source flux to the total flux, including the unmagnified source flux and flux from any nearby sources blended in the source's aperture. 

Since the planet has to be close to one of the images of the primary event in order to perturb the image, we can assume that the planet has separation $s$ equal to the distance of the image from the primary lens which is given by Equation \ref{eqn:find_s} where $t_p$ is the time of the planetary perturbation \citep[e.g.][]{2012ARA&A..50..411G} .
For that reason, we can calculate $s$ using this equation, along with the measurements of $u_0$ and $t_p$ from the light curve:
\begin{equation}\label{eqn:find_s}
    s = \frac{u \pm \sqrt{{u}^2 + 4}}{2} ,\; {\rm where} \; u = \sqrt{{u_0}^2 + {(\frac{t_0 - t_p}{t_E})}^2}
\end{equation}
It is important to note that this approximation is expected to work for planetary perturbations caused by the planetary caustic, since this method is in fact finding the separation between the caustic and the lens star, and only the planetary caustic separation from the lens star can be a reasonable representation of star/planet separation. 

If the source star's apparent size is small compared to the planet's caustic, the duration of the planetary perturbation is equal to the time it takes for the image of the single-lens event to pass by the position of the planet and can be approximated by $t_{Ep} = t_E q^{1/2}$. However, in cases where the source projected size is not small, it affects the duration of the planetary event resulting in a continuous degeneracy between mass ratio and finite source size. In this regime, the duration of the planetary event is determined both by mass ratio $q$ and source angular size $\rho$ which is equal to the ratio of source crossing time, $t_\ast$ to the Einstein timescale, $t_E$ \citep{gaudi1997planet}. Therefore, in this model we cannot directly find the mass ratio by measuring the duration of the planetary perturbation. We measure $t_{Ep}$ from the light curve and compare that to the true value of $t_{Ep}$ obtained by the quadrature sum of $q$ and $\rho$ as in the equation,
\begin{equation}\label{eqn:find_q}
    {t_{Ep}} = \sqrt{(q + {\rho}^2)} \times{t_{E}} ,\; \rho = \frac{t_{\ast}}{t_E} 
\end{equation}

In the next subsection, we will introduce our algorithm and explain how we have implemented this model in the algorithm to achieve the goals of this approach.

\subsection{The algorithm}\label{subsec:algorithm}

The algorithm we employ first attempts to fit a single-lens function to each lensing event.  We then subtract the best-fit single lens model, and search for signs of a binary lens (indicated by a secondary event) in the residuals.

In the first step, we fit a Point-Source-Point-Lens (PSPL) function (first term in Equation \ref{eqn:A-2}) to the data. We first select a set of initial guesses for the four PSPL parameters $t_0$, $t_E$, $u_0$ and $f_s$, and then minimize the ${\chi}^2$ of the PSPL functional fit to the data. The initial guesses of those four variables are the algorithm input parameters, and we employ an automated method to generate reasonable values for them.
 
We find the maximum differential flux $F_{max}$ by selecting the maximum measured value of $F(t)$, and setting the time of maximum flux as the initial guess for $t_0$. Then, we assume $A_{max} \simeq \frac{1}{u_0}$, and rearrange Equation \ref{eqn:fs-1} to estimate the initial guess for $u_0$. 
\begin{equation}\label{eqn:fs-2}
    u_0 \simeq \frac{f_s}{F_{Max}+f_s+1}
\end{equation}
The parameter $f_s$ is the ratio of the unmagnified source flux to the baseline aperture flux, so it should be fitted as a part of the PSPL function. In reality, one would do a more extensive initial parameter search to find initial guesses for $f_s$ and $u_0$, but in order to focus our efforts on the anomaly fitting rather than finding initial guesses for the single-lens fit, we assumed that a full search of the single lens parameter space would be certain to find a model consistent with the true blending parameter. We then adopt the true value of $f_s$ as the initial guess and the initial guess for $u_0$ is obtained from Equation \ref{eqn:fs-2}.

We attempt the PSPL fit with two different initial guesses for $t_E$. The first one assumes that $u_0=0$, such that $t_E$ is equal to the time between the closest approach and when the source reaches the Einstein ring ($u=1$). The second one assumes that $t_E$ is equal to the full-width at half-maximum (FWHM) of the primary event. We then accept the fit that has a lower value of $\chi^2$.

After we fit the single-lens model, we use a smoothing algorithm to reduce the noise in the residual light curve. The smoothing algorithm includes a moving box of $10$ data points that slides over the light curve and convolves onto it. Therefore each data point in the light curve is replaced by the average of itself along with its next $9$ data points. This will allow the algorithm to detect the outliers more easily. The algorithm then searches for a maximum or minimum point in the residuals, whichever is greater in magnitude. Assuming that the peak or trough is caused by a secondary microlensing event due to a planet around the lens star, we fit the second function (Equation \ref{eqn:A-2}) to the data to describe the planetary perturbation. Thus, in addition to the parameters from the first fit we add three more initial guesses for the planet. 

These three initial guesses for the planet will be also found automatically after the first fit. So the time of the largest peak or trough in the residuals of the first fit will be assigned as $t_p$, and its height as $a$. For the Einstein timescale of the planetary event, we test values of 0.01, 0.1 and 1 days, and then selects the value that results in the smaller $\chi^2$.

\begin{equation}\label{eqn:chi_2}
    {\chi}^2 = \frac{\sum {(f-x_i)}^2}{{{\sigma}_i}^2}
\end{equation}

For fitting each of the functions, we use the ``Nelder Mead" or downhill simplex method of minimizing the ${\chi}^2$ function \citep{nelder1965simplex}. Equation \ref{eqn:chi_2} shows the ${\chi}^2$ function we have used, where $f$ is the model (first term in Equation \ref{eqn:A-2} in the two terms in Equation \ref{eqn:A-2} in the second fit) and $x_i$ and ${\sigma}_i$ are the data and the error in data, respectively. We refer to ${{\chi}_1}^2$ as the ${\chi}^2$ of the PSPL fit and ${{\chi}_2}^2$ as the ${\chi}^2$ of the PSPL+Gaussian fit, and the difference between these two as ${\Delta \chi}^2$. 

\begin{figure}[H]
    \centerline{\includegraphics[width=11cm]{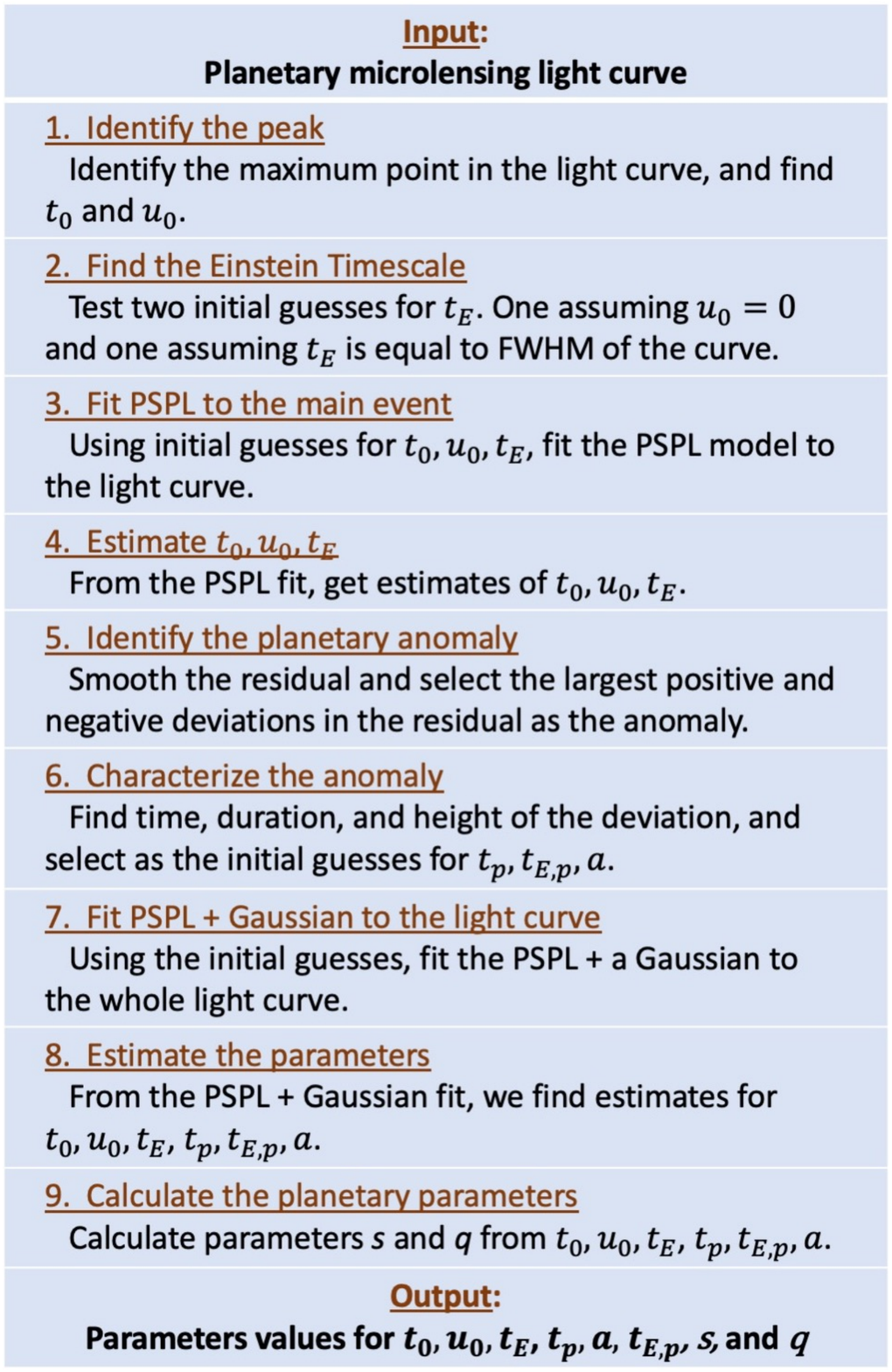}}
    \caption{Flow chart of the steps taken by the algorithm from receiving the data to finding the final parameters.}
    \label{fig:flowchart}
\end{figure}

In the next step, the results of the fits are put in Equations \ref{eqn:find_q} and \ref{eqn:find_s} to estimate $s$ and $q$ for each light curve. We see in Equation \ref{eqn:find_q} that the mass ratio is related to the square root of the duration of the planetary event, which means smaller planets have shorter events. Projected separation, calculated from Equation \ref{eqn:find_s}, can have two values. This is where the algorithm decides if the planet lies inside the Einstein Ring $(s<1)$ or outside of it $(s>1)$. We check if the planetary perturbation was found due to a peak or a trough in the first place. If the perturbation is detected is as a peak then the value of $s$ larger than unity is chosen and vice versa for the trough. After calculating $s$ and $t_{Ep}$ from the fits, we compare them to their true values for the analysis in the next section. In the flow chart of Figure \ref{fig:flowchart} we have summarized the steps taken by the algorithm from receiving the data to generating the results.

In the next step, we apply our algorithm to all light curves, and then label them as successful or failed recoveries based on the parameters ${\chi}^2$ and $u_0$, with numerical cuts as described in the next section.

\section{Simulated Data}\label{sec:sim_data}

To assess the effectiveness of our algorithm we have used a set of simulated microlensing light curves generated for the WFIRST mission. Full details of how the light curves were generated is given by \citet{penny2013exels} and \citet{Penny2018Predictions} (hereafter P13 and P18); here we only describe the details relevant to our tests. The simulated light curves are based on the WFIRST AFTA mission design, and have six 72-day long seasons spread through the WFIRST mission, observed with 15 minute cadence. 

The simulations output light curves if the combined $\Delta \chi^2$ of a single lens fit relative to the true underlying model used to generate the event was greater than 60 for all three observatory designs combined. This threshold results in an effective $\Delta \chi^2$ threshold for a single observatory that is usually much less than 60. For comparison, P18 used a $\Delta\chi^2>160$ threshold to define a planet detection, so we expect the set of light curves to contain many planet signals that are only marginally detectable. 

The parameters of the simulated light curves were drawn uniformly from the ranges $u_0=[-3,3]$ and $t_0=[0,2010]$ (note only 432 of these days contained WFIRST observations, and there was a large gap of $\sim$840 days between two groups of three seasons and gaps of 111 days between the seasons). The lens and source properties that determined the event timescale, source magnitudes and blending were drawn from the Besan{\c c}on population synthesis Galactic model \citep{kerins2009synthetic} as described in P13 and P18. Planet semimajor axes were drawn from a log uniform distribution in the range $0.3<=a<30$ AU, and masses were drawn from the fiducial planet mass function described in P18, a power-law rising towards low masses before saturating at ${\sim}5.5 M_{\oplus}$, based on the \citet{2012Natur.481..167C} measurement.

These microlensing events have Einstein timescales between $0.2$ and $>1000$ days with a median of 16 days. The mass ratio $q$ is between $10^{-7}$ and $0.3$ with a median of $10^{-4}$. The projected separation $s$ is between $0.01$ and $60$ Einstein radii with a median of $1$. The events have impact parameters $u_0$ ranging between $10^{-6}$ and $3$ with a median of $0.3$ in units of the Einstein radius.

\section{Results}\label{sec:results}

We apply the algorithm described in Section \ref{subsec:algorithm} to all 13,206 simulated light curves. We then identify failure modes in the algorithm and set up conditions to exclude the failed ones. 
After that, we compare each of the fitted parameters to the true values to assess the performance of the algorithm. The three parameters $t_0$, $t_E$, and $u_0$ shared with the single-lens model are mostly fitted easily and the final parameters have very small deviation from the true ones. We then focus on the two parameters describing the planetary anomaly, $s$ and $q$, that we calculate using the planetary parameters $t_p$, $t_{Ep}$, and $a$, that we get from fitting the anomaly. Since finite source effects are present, we cannot measure $q$ directly using the planetary perturbation duration, but we measure $q+{\rho}^2$ instead (see Equation \ref{eqn:find_q}). 

\subsection{Initial sample cuts}\label{subsec:initial cuts}

\begin{figure}[H]
    \centerline{\includegraphics[width=14.5cm]{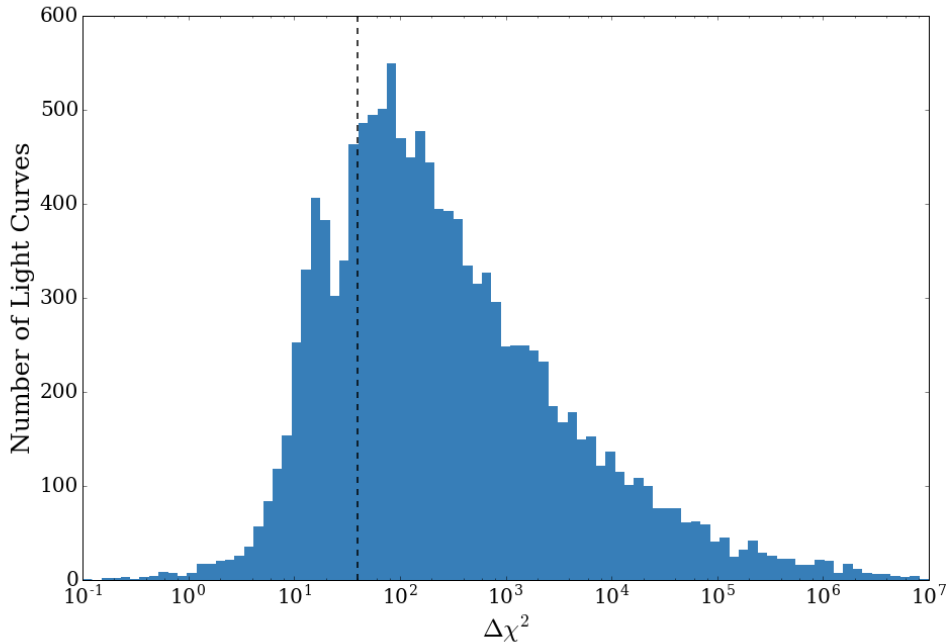}}
    \caption{Histogram of ${\Delta \chi}^2$, the difference in ${\chi}^2$ between the single lens fit and one with a planetary perturbation. Two distributions can be seen in the plot. The black dashed line shows the cut-off where events with ${\Delta \chi}^2<40$ have been excluded.}
    \label{fig:hist_chi}
\end{figure}

We perform the fits as described in section \ref{subsec:algorithm} for the 13,206 light curves, with no prior information about the degree of coverage of the main event and the planetary event.  There are several situations in which we do not expect an automated method to successfully measure the planetary parameters:
\begin{itemize}
    \item If the amplitude of the anomaly is much smaller than the noise in the data, the planetary perturbation will not be detectable by our algorithm.
    \item If the planetary event happens to have multiple peaks in the light curve, which can be caused by caustic crossings, the PSPL+Gaussian model will not fit the light curve well.
    \item If the planetary event is caused by the central caustic in a wide or close topology, the algorithm finds the anomaly close to the peak of the main event, and then deduces that the system has projected separation close to one where in reality, the planet is located outside or inside the Einstein ring.
\end{itemize}
We identify these cases and eliminate them from the tested sample so as to examine the performance of our approach in the cases where we intend for it to work.  In order to exclude the first failure mode above, we require that ${\Delta \chi}^2$, the difference between the ${\chi}^2$ of the first fit (PSPL) and the ${\chi}^2$ of the second fit (PSPL+Gaussian), to be larger than $40$ to ensure detectability and exclude the events with no distinguishable planetary signature. As mentioned in section \ref{sec:sim_data}, P18 used a threshold of 160 for declaring a planet to be detectable. When we adopted this threshold, the improvement in parameter estimate precision was modest, so we concluded that our algorithm is effective even for planetary signals with low significance. Figure \ref{fig:hist_chi} shows the overall distribution of the values of ${\Delta \chi}^2$, in which two populations are apparent.  

We conducted visual checks of a sample of the light curves with ${\Delta \chi}^2 <40$ and ${\Delta \chi}^2 >40$, and we found that the threshold of 40 successfully selects cases where the addition of the Gaussian component significantly improves the fit, and excludes failed fits. This cut excludes $23.3\%$ of total events. 

We do not expect our algorithm to work on multi-peaked events such as those with caustic crossings. To identify such events, we compute the $z$-score of each light curve \citep{kreyszig2010advanced} using a code implemented in python\footnote{https://gist.github.com/ximeg/587011a65d05f067a29ce9c22894d1d2}.
We then classified events as multi-peaked if they had more than two data points with {\it z}-scores larger than 6. The {\it z}-score for each data point was calculated as $(x-\mu )/ \sigma$ within the previous 30 data points. This cut excludes $5.4\%$ of all events.

In central caustic events, when there is no planetary caustic crossing, there is no large planetary signal close to the time when the source is close to the planet. The algorithm then detects the signal close to the peak of the stellar lensing event, making $|t_0 - t_p|$ very small and therefore $u$ (see Eqn \ref{eqn:find_s}) becomes small and $s$ is found to be incorrectly close to unity. We therefore added another criterion that requires the fitted value of $u_0$ to be larger than 0.045 Einstein radii, which excludes $23.7\%$ of the total events. After applying this condition, 7,108 light curves remain, with a total of $45.4\%$ light curves excluded by these cuts. 

We then exclude the light curves that have $s_{fitted}>5$, because a few events with $s>5$ will have both the planetary anomaly and the stellar event in a season and also these events can be examined separately as candidates for free floating planets. This cut will exclude $7.2\%$ of total events.

\begin{figure}[H]
    \centerline{\includegraphics[width=14cm]{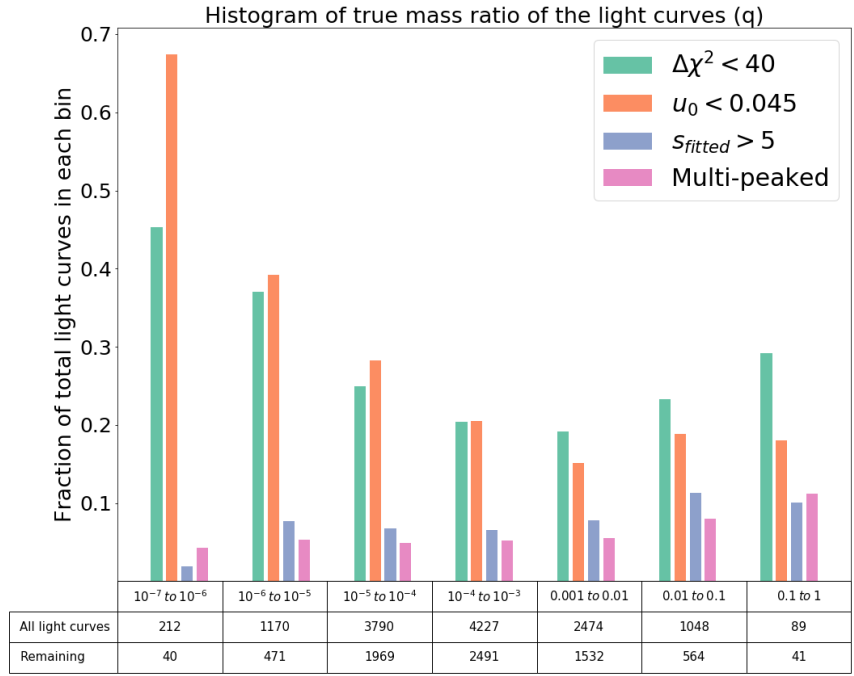}}
    \caption{Fraction of light curves eliminated by each cut across ranges of mass ratio. The table below the histogram shows the number of total light curves in each bin along with the number of the light curves surviving the cuts.} 
    \label{fig:hist_q_cuts}
\end{figure}

\begin{figure}[H]
    \centerline{\includegraphics[width=14cm]{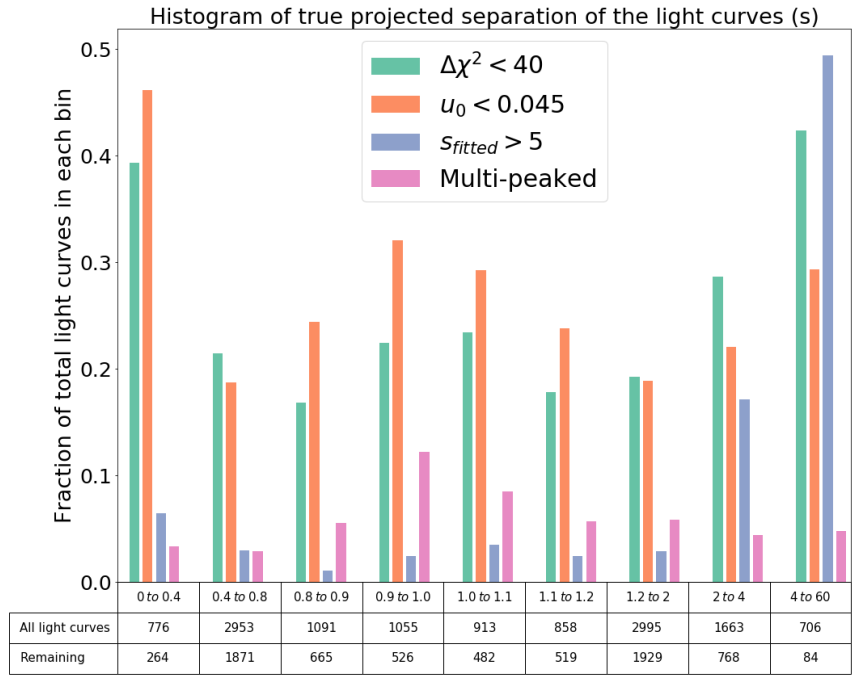}}
    \caption{Fraction of light curves eliminated by each cut across ranges of projected separation. The table below the histogram shows the number of total light curves in each bin along with the number of the light curves surviving the cuts.}
    \label{fig:hist_s_cuts}
\end{figure}

The histograms in Figures \ref{fig:hist_q_cuts} and \ref{fig:hist_s_cuts} show distribution of parameters $q$ and $s$ of the light curves excluded by the cuts above. The plots show the fraction of the light curves excluded by each cut, in each parameter bin. For example, in Figure \ref{fig:hist_q_cuts}, out of all light curves with mass ratios between 0.1 and 1, $30\%$ were excluded by the $\Delta{\chi}^2$ cut, $18\%$ were excluded by the $u_0<0.045$ cut, $11\%$ were excluded by the $s_{fitted}<5$ cut, and $10\%$ were excluded by the multi-peaked cut. Note that a given light curve can fail more than one cut, so the fractions do not sum to unity in each bin. A table below each histogram shows the number of total light curves in each bin along with the number of the ones that survived the cuts These plots help us understand the results of the cuts as a function of mass ratio, and projected separation. While we have described the excluded events as failures, it is better to think of them as being events belonging to a different class that requires a different method to be applied.


\subsection{Examples of approximate model fits}\label{subsec:examples}
    
\begin{figure}[H]
 \centerline{\includegraphics[width=11cm]{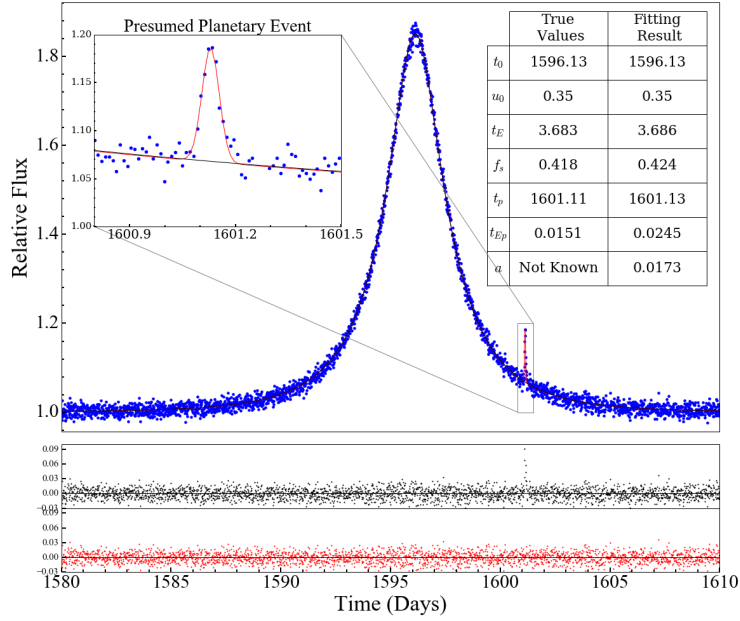}} 

    \caption{An example of a good fit between the model and the data for a system with projected separation larger than unity. The vertical axis represents the relative flux of the source, and the horizontal axis is time. The top panel shows the simulated light curve (blue data points), the PSPL model (black curve) and the fitted PSPL+Gaussian model (red line), and the inset is a zoom-in of the planetary event. The two bottom panels show the residuals of the PSPL and PSPL+Gaussian fit.}
    \label{fig:good-fit1}
 \end{figure}

\begin{figure}[H]
  \centerline{\includegraphics[width=11cm]{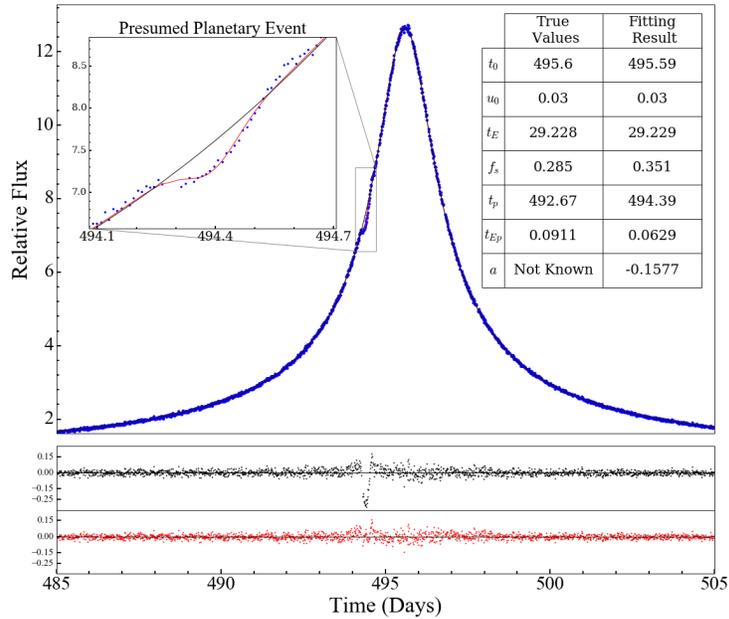}}
  \caption{ An example of a good fit between the model and the data for a perturbation in minor image. See Figure \ref{fig:good-fit1}t for a description of the plot features.}
  \label{fig:good-fit2}
\end{figure}

To give the reader an idea of how our algorithm performs on simulated data, in this section, we examine a few examples. In each example plot, a table shows the fitted and true parameters. The red data points in the residuals (bottom panel) are relative to the PSPL+Gaussian model, and the black data points are relative to the PSPL model. A good fit would result in removing the features in the black residual and therefore having an almost smooth red residual. In Figures \ref{fig:good-fit1} and \ref{fig:good-fit2}, the PSPL+Gaussian model (red curve) follows the data (blue data point) very well whereas the PSPL model (black curve) does not describe the anomaly. The table also shows that the fitted parameters are very close to the true values. Note that the fit in these light curves that we consider as successful are not perfect, but they do capture the salient features of the events (peak height, duration, etc.). 


\begin{figure}[H]
  \centerline{\includegraphics[width=11cm]{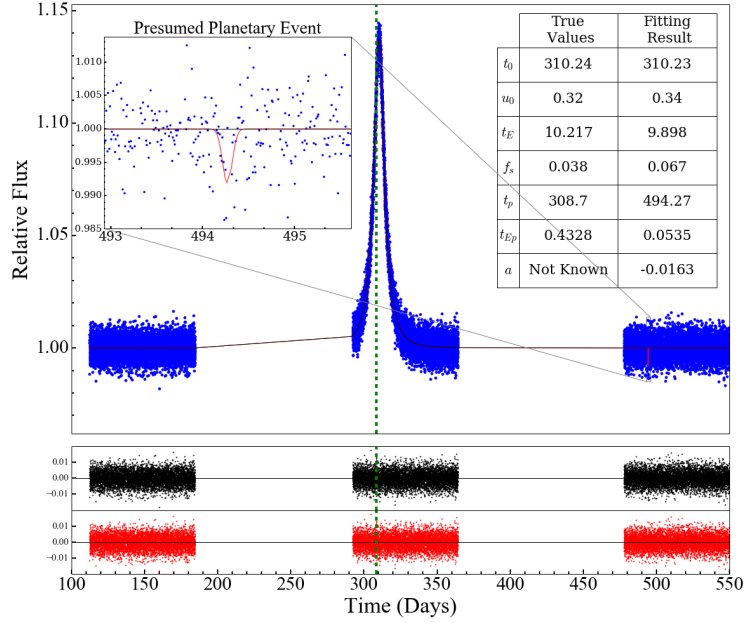}}
  \caption{ An example of a poor fit between the model and the data where the algorithm has chosen outlying data points at time 494 as the planetary signal whereas the true anomaly is shown with the green dashed line. Because the amplitude of the perturbation is small compared to the noise, even after smoothing the residual, the planetary event could not be detected.}
  \label{fig:bad-fit1}
\end{figure}

\begin{figure}[H]
  \centerline{\includegraphics[width=11cm]{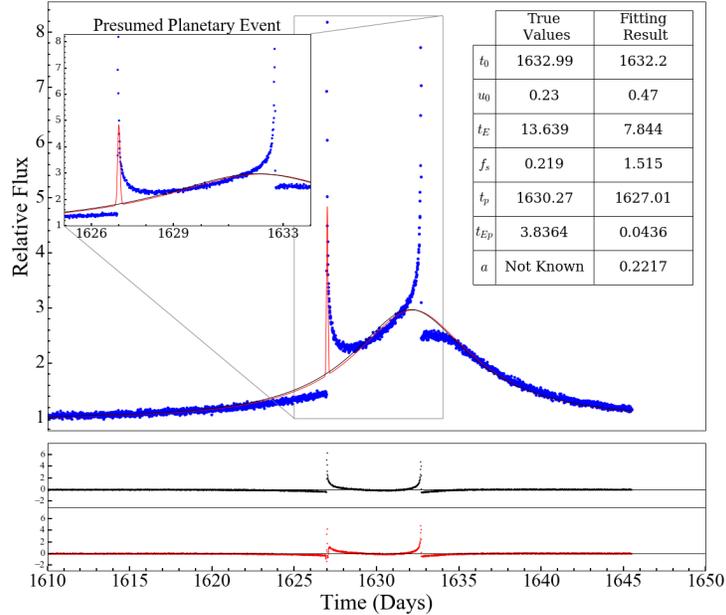}}
  \caption{  An example of a poor fit between the model and the data where there are multiple caustic crossings and the model cannot fully describe the data.}
  \label{fig:bad-fit2}
\end{figure}

Looking at Figure \ref{fig:bad-fit1}, we can see the algorithm has chosen a set of outlier data points to be the planetary detection, whereas the real planetary perturbation is impossible to see because of its small fractional deviation from the single-lens model. This is an example of a target where the PSPL+Gaussian fit is very similar to the PSPL fit ($\Delta{\chi}^2<40$), such that the planetary signal is not large enough to be detectable. Figure \ref{fig:bad-fit2} shows another example of the failure of our algorithm, and that happens when there is multiple caustic crossings and our model cannot fully describe the data. We used a z-score algorithm for detecting these events and excluding them. Note that in some of these multi-peaked events the time of the planetary anomaly has been estimated correctly which is not very unexpected since $X_c$ in Equation \ref{eqn:find_s} is an estimate of the caustic location. Additionally, an alternative, non-parametric estimate of the anomaly duration could be used to get a better estimate of $q$ in these situations.


\begin{figure}[H]
  \centerline{\includegraphics[width=11cm]{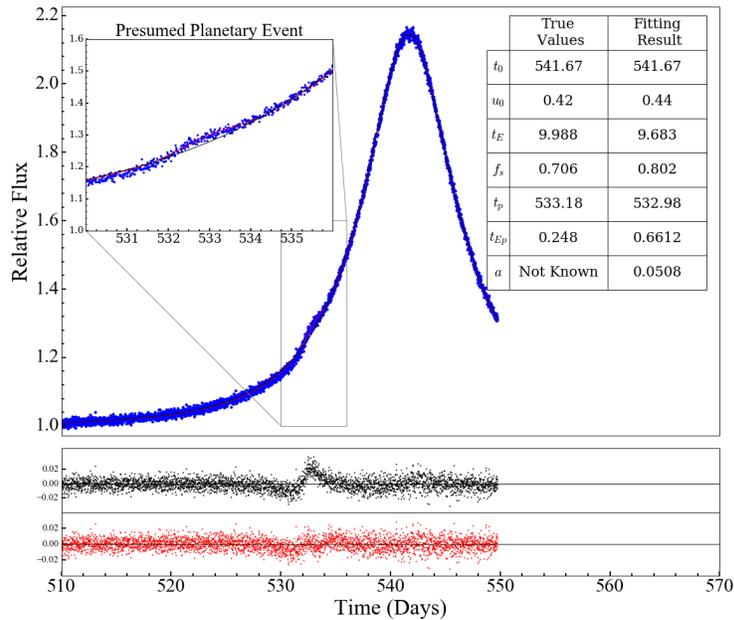}}
  \caption{ An example of the discrete degeneracy in $s$.  In this case, the algorithm has correctly identified the time of the planetary perturbation, but has found the peak of the anomaly to be larger than its trough, and therefore incorrectly assumed this is a major image with $s>1$.}
  \label{fig:bad-fit3}
\end{figure}

\begin{figure}[H]
  \centerline{\includegraphics[width=11cm]{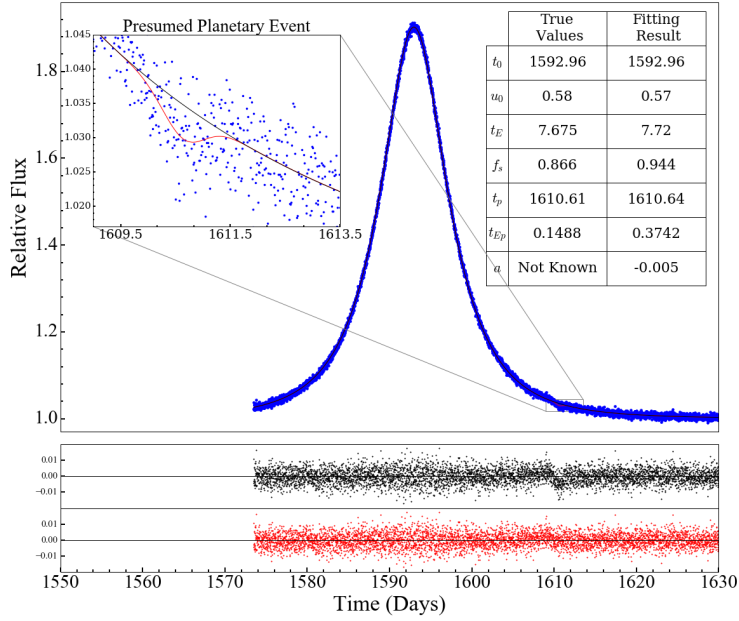}}
  \caption{Another example of the discrete degeneracy in $s$ (similar to Figure \ref{fig:bad-fit3}) where the algorithm has found the trough of the anomaly to be larger than its peak, and therefore incorrectly assumed this is a minor image with $s<1$.}
  \label{fig:bad-fit4}
\end{figure}

Figures \ref{fig:bad-fit3} and \ref{fig:bad-fit4} are two examples of the discrete degeneracy in $s$, where the algorithm has incorrectly chosen $1/s$ as the projected separation of the system. In Figure \ref{fig:bad-fit3}, we have a minor image, but the caustic crossing for these events include both peaks and troughs, and peaks can be higher in amplitude than the troughs and therefore in these cases the algorithm cannot correctly resolve the degeneracy. In Figure \ref{fig:bad-fit4}, we have a major image with a very small signal, and in these cases the algorithm fails to see that the peaks are higher than the troughs. These are therefore not complete failures, and could potentially be identified by additional automated filtering. In Section \ref{sec:results}, we show how manually addressing these cases can help improve the overall results.


\subsection{Performance of the algorithm}\label{subsec:performance}

\begin{figure}[ht]
     \centerline{\includegraphics[width=15cm]{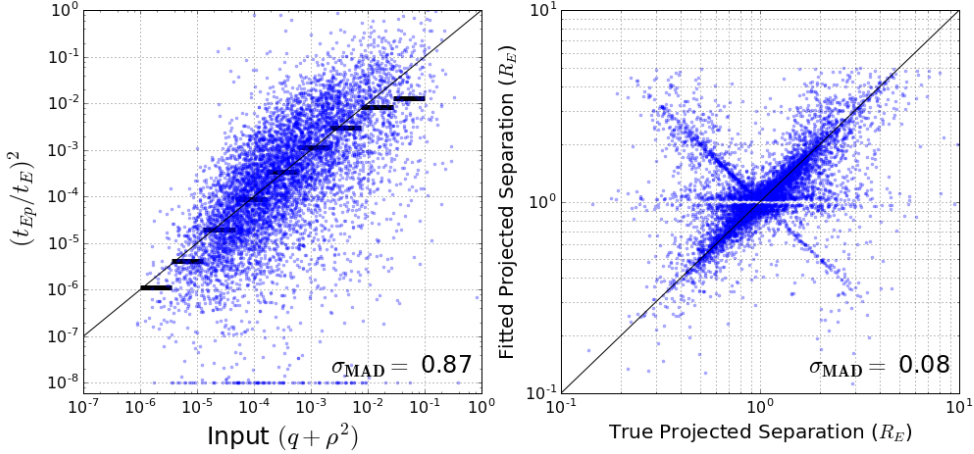}  }
     \caption{Log-log plots of fitted physical parameters versus the true values for the light curves after applying the initial cuts. {\it Left:} Fitted value versus the true value of ${(t_{Ep}/t_E)}^2$. Black bars are binned medians of the plot showing the overall trend. The algorithm is more successful at middle range values of $q+{\rho}^2$, and greater deviation from the 1-1 line is seen for higher and lower ranges of $q+{\rho}^2$. Note that some of the data points on this plot have value of ${(t_{Ep}/t_E)}^2$ lower than $10^{-8}$, and we show them on line $y=10^{-8}$ for a better visualization. {\it Right:} Fitted value versus the true value of projected separation in units of the Einstein radius. 
     }
     \label{fig:res-1}
 \end{figure}

Figure \ref{fig:res-1} shows plots of the fitted versus true values for the parameters $s$ and ${(t_{Ep}/t_E)}^2$ for the 7,108 light curves that pass the initial cuts described above. Each data point in this plot represents the result of the fit to one light curve. Note that we have selected the parameter ${(t_{Ep}/t_E)}^2$ instead of $q$ because the approximation $q = {(t_{Ep}/t_E)}^2$ is only valid when the finite source effect is negligible compared to the duration of the planetary anomaly. Here, we compare ${(t_{Ep}/t_E)}^2$ with $q+{\rho}^2$ to account for the impacts of the finite source effect as described in Equation \ref{eqn:find_q}. We have calculated Median Absolute Deviation (MAD) as a measure of scatter in our plots in this paper. It is worth mentioning that since our simulated events were drawn to pair up any two stars as long as the source was further away than the lens, an over-representation of events with extreme timescales and small Einstein radii could exist in the dataset. In order to account for that we also calculated the weighted MAD of the results where weights are $\gamma = \mu_{rel} \times \theta_E$; $\mu_{rel}$ is the relative proper motion and $\theta_E$ is the angular Einstein radius. The weighted MAD was not significantly different from the unweighted MAD, and therefore we do not incorporate the weights in the rest of the paper. Also, instead of reporting the MAD itself, we report $\sigma_{MAD} = \sim1.48\times MAD$, an estimate of the standard deviation, which we refer to henceforth as the ``scatter".

In the plot of fitted versus true projected separation (right panel), most of the fitted parameter values are fairly close to the true values, but there is substantial scatter. We find a scatter of $\sigma_{MAD}=0.08$ for the projected separation of all 7,108 light curves in logarithmic units, showing that this method is able to determine log of separations to about 8\% precision or in other words, has a fractional precision of $\sim 18\%$ on $s$.

One can see several patterns in the plot of projected separations.  There is a set of data points on the $x=-y$ line; those are cases where the fitted separations are the inverse of the true separations.  As mentioned before, that is due to a common discrete degeneracy in microlensing, where $s$ and $1/s$ are obtained from the same set of parameters (see Equation \ref{eqn:find_s}). For these cases, the value of $X_C$ is correctly estimated but the algorithm has failed to recognize which of the peaks or troughs in the residuals had a larger amplitude. Another feature in the plot is the small horizontal gap in points at $s_{fitted}=1$, which is due to the requirement that $u_0 > 0.045$ as mentioned before. 
 
The left panel of Figure \ref{fig:res-1} shows the fitted versus true values of ${(t_{Ep}/t_E)}^2$, our approximation for the mass ratio.  The algorithm is able to recover planetary duration to within one order of magnitude for most of the light curves, and the  logarithmic scatter of ${(t_{Ep}/t_E)}^2$ relative to $q+{\rho}^2$ is $0.87$ implying that the fractional precision on $q+{\rho}^2$ is $\sim 200\%$. The binned median is also shown on this plot (black horizontal bars) and indicates the trend of the data points. It shows that in estimating the value of $q+{\rho}^2$, our algorithm has been more successful at intermediate mass ratios. Comparing this plot with the projected separation plot (right panel) shows that our method is more successful for estimating projected separation rather than planetary event duration. The reason for that is that we find the duration of planetary perturbation from the light curve, and the duration depends on the geometry of the source trajectory and size of the caustics. When planets are smaller, and the finite source effect is not negligible, the duration of the perturbation is determined by both mass ratio and finite source effect (see Equation \ref{eqn:find_q}). Additionally, it seems that the estimated perturbation duration tends to be smaller than the true value, and that happens because the estimated duration of the planetary perturbation is determined by the width of the fitted Gaussian curve, which tends to underestimate the true duration of the planetary events.

\begin{figure}[H]
     \centerline{\includegraphics[width=11cm]{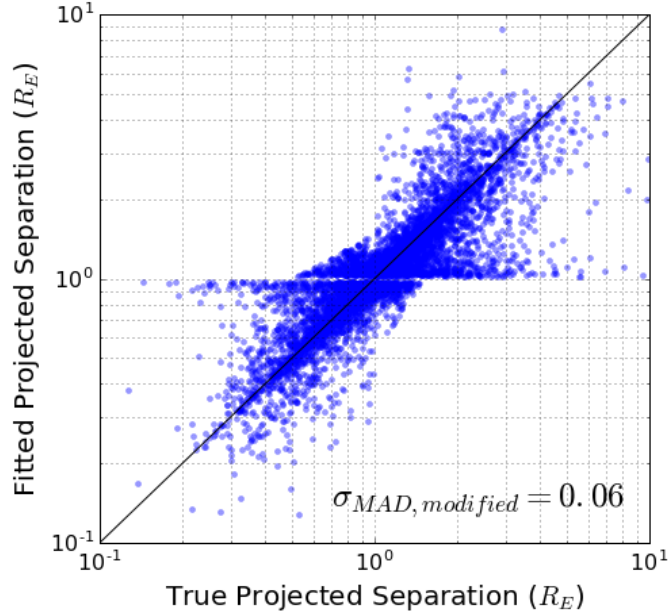}  }
     \caption{Log-log plot of the fitted value versus the true value of projected separation in logarithmic units. In this plot, degenerate values of $s$ have been manually changed to $1/s$ assuming that this degeneracy can be resolved in later analysis. This modification decreases the scatter by 0.02. 
     }
     \label{fig:res-2}
 \end{figure}

Because of the possibility for both positive and negative anomalies due to planetary caustics, we expect that any detailed light curve modelling would check both the $s>1$ and $s<1$ solutions for the planet position. Therefore, in Figure \ref{fig:res-2} we re-plot the projected separation $s$.  In this case, $s_{fitted}$ is chosen to be either the algorithm-derived value or its reciprocal, such that the event resides in either the upper right quadrant ($s_{true}>1$ \& $s_{fitted}>1$) or the lower left quadrant ($s_{true}<1$ \& $s_{fitted}<1$).  That is, we assume that events that suffer from the $s$ degeneracy will be properly identified with this algorithm, and the degeneracy properly resolved with later analysis. We then recompute the scatter and obtain a significantly smaller value of $\sigma_{MAD}=0.06$

\begin{figure}[H]
     \centerline{\includegraphics[width=13cm]{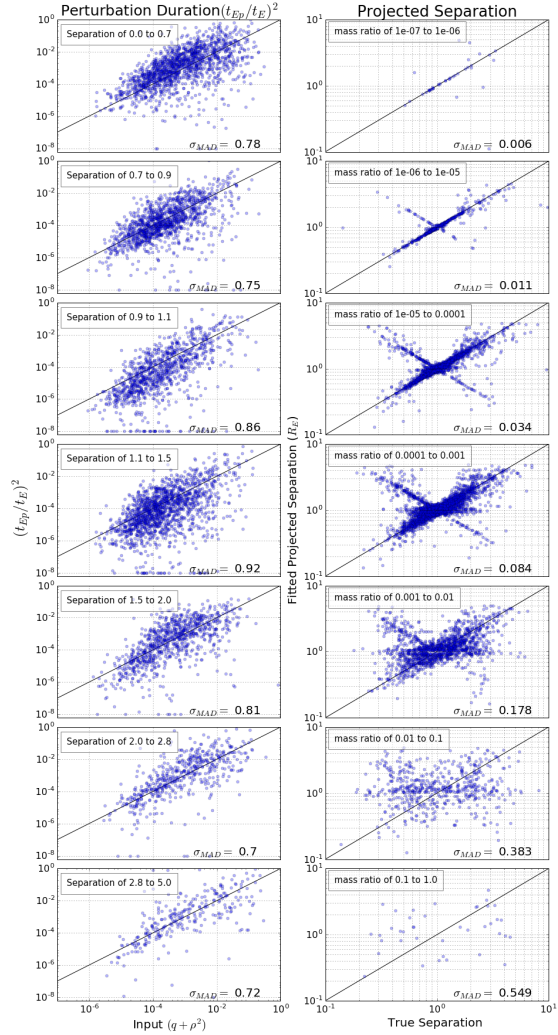}  }
     \caption{ Plots of the fitted parameters $s$ and ${(t_{Ep}/t_E)}^2$ against their true parameters in different ranges of mass ratios and projected separations. {\it Left column:} Fitted ${(t_{Ep}/t_E)}^2$ versus the true value ($q+{\rho}^2$ according to the Equation \ref{eqn:find_q}) in logarithmic scale in different ranges of projected separation. The overall $\sigma_{MAD}$ does not change significantly in different ranges and most of the fitted values are within one order of magnitude of the true values. {\it Left column:} Fitted $s$ versus the true value in logarithmic scale in different ranges of mass ratios. The scatter decreases when going from higher mass ratios to lower ones showing the higher rate of success of the algorithm for wide and close topologies.}
     \label{fig:appc-all}
 \end{figure}

We also investigate how successful this algorithm performs in different ranges of mass ratios and projected separations, as plotted in Figure \ref{fig:appc-all}. The right column displays plots of the fitted versus true values for planet-star projected separation in logarithmic units in seven ranges of true mass ratios. We observe that the scatter decreases significantly for lower mass ratios. The reason is that as mentioned earlier, we expect this method to work for planetary perturbations caused by planetary caustics in wide and close topologies. At lower mass ratios, only a small regime of separations result in intermediate topology \citep{gaudi2010exoplanetary}, and therefore we expect most of the lower mass systems to be in wide/close topologies resulting in an improvement in the reliability of the fitted values of $s$ from our method. At higher mass ratios, a large regime of $s$ lies in the intermediate topology region and therefore the planetary event is caused by the central caustic the resulting fitted values of $s$ are not reliable.

The left column of Figure \ref{fig:appc-all} also shows the fitted values versus true values of ${(t_{Ep}/t_E)}^2$ in logarithmic units in different ranges of planet-star projected separations. The overall scatter in these plots does not change drastically in different regimes of projected separations. Overall, we can say that for $\sim 70\%$ of the light curves we are able to estimate the value of ${(t_{Ep}/t_E)}^2$ to within one order of magnitude.

\section{Conclusion}\label{sec:conclusion}

\subsection{Summary}

We have investigated the effectiveness of the \citet{gaudi1997planet} ``by-eye" fitting method for planetary microlensing events. To do this, we have developed an algorithm to encode the critical features of the method, such that it could work unsupervised and with no input other than light curve data. To test the algorithm, we applied it to simulated WFIRST light curves and we find that our algorithm successfully recovers the physical parameters for certain types of microlensing events. Our results demonstrate quantitatively for the first time that the \citet{gaudi1997planet} method for estimating planet parameters is a reasonably accurate method, with a fractional logarithmic precision in $s$ of  $\sim $ 6\% and a fractional logarithmic precision in $q$ of  $\sim $ 87\%.  The method is subject to the $s$ vs. $1/s$ degeneracy, and is less successful for high mass ratio events, and events caused by the central caustic. In other words, this algorithm has a higher success rate for events with wide and close caustic topologies, and when the caustics are smaller. 

The algorithm therefore allows an automated quick search of the entire microlensing binary-lens data set for lower mass planets. The scatter for the projected separation in logarithmic scale in the mass ratio range from $10^{-6}$ to  $10^{-5}$ is $0.007$, whereas the scatter in the range of mass ratio from $0.001$ to $0.01$ is $0.12$. The difference in scatter for the mass ratio plots in different ranges of projected separation is less significant, but still shows improvement. For projected separations of $0.8$ to $0.9$ Einstein radii, the scatter for logarithm of mass ratio is $0.48$, while for projected separations of $1.1$ to $1.2$ Einstein radii, it is $0.71$. It is also useful to note that the average time of running the code for each light curve is about $50$ seconds on a 2.7 GHz Intel Core i5 iMac computer with $16 \; GB$ RAM for light curves with $40K$ data points.

The algorithm that we have presented performs reasonably well at estimating the projected separation and mass ratio for a subset of planetary microlensing events. We do not expect that these estimates would be treated as robust measurements of the quantities of interest, but as valuable heuristic estimates. In this way the algorithm could be used to identify candidate planetary microlensing events and provide initial guesses for full light curve modeling analyses. This could be useful for quickly identifying low-mass planet candidates for follow-up by observatories from the ground or space, potentially enabling measurement of microlensing satellite parallax \citep[e.g.][]{gould1992discovering,zhu2015spitzer} or other quantities that require observations during an event. One avenue for future work would be to investigate how well downhill fitting performs on events where the algorithm has estimated the initial parameter guesses.

\subsection{Further Improvements}

There are multiple ways to improve this method. While the algorithm only works on a subset of planetary microlensing events, its fitted parameters could conceivably be used as inputs, along with other easily measurable quantities, into machine learning classifiers. In fact, in addition to the two statistics $X_C$ and $t_{Ep}/t_E$ that we compute here, minor extensions to the algorithm could be used to cheaply compute a wide range of quantities that would carry some information about the planet. 

The simulated data here consisted of only binary-lens events. By adding other types of microlensing events and also other variabilities, our dataset can become more similar to what WFIRST will actually observe.  This method should be able to distinguish single-lens events from binary-lens events including planets, at least for the cases where the planetary event is large enough to measure the planet properties. We have assumed that the task of reliably classifying variables as microlensing will be performed by another algorithm \citep[e.g.][]{kim2018korea}.

There are currently a number of methods that attempt to categorize photometric variability in large data sets, such as statistical methods and machine learning. Machine learning methods for detecting different types of variabilities is becoming more common \citep[e.g.][]{richards2011machine,pichara2013automatic,pashchenko2017machine,valenzuela2017unsupervised}. For example, \citet{belokurov2003light} uses neural networks to detect microlensing light curves among variety of variabilities by trying to select the non-periodic symmetric variabilities. This approach can be implemented in the code, so that it would be able to first detect microlensing events or even possibly categorize them as single-lens and binary-lens or multiple-lens events.
 
It is also useful to investigate how astrophysical variability in the source star could affect these results. Uniform photometric surveys like Kepler can provide the rate of various types of variability for certain stellar populations, which can then be added to the simulated light curves.  We can then test if this algorithm can still extract the planetary parameters with a similar scatter.  We should also consider cases where there is variability in the stars close in angular separation to the source stars. In that case, the variability will not be magnified, but the non-magnified the light curve will show the variability. 
Additionally, we can see whether including orbital parallaxes in the simulated data affect the performance of the algorithm. Measuring this parallax helps constrain masses and the distance to the lens host star and planet. Additionally, the algorithm presented here was used on full light curves, and it would be important to show it works on partial light curves as well.

\subsection{Discussion}

We have presented a fast approach for analyzing binary-lens microlensing events. This method, unlike traditional approaches that require significant time and computational power, attempts to characterize planetary signatures in microlensing light curves using simple functional fits. We have applied this method to a sample of simulated WFIRST binary-lens microlensing events, and conclude that this method can work fast and more efficiently for planetary systems with smaller planet-star mass ratios and planet-star projected separations larger than the Einstein radius $(s>1 \; or \; s<1)$. In other words, for systems with wide and close topologies, this approach is more likely to succeed.  We intend to pursue and refine this approach using a more complete data set containing additional types of stellar photometric variability to improve our method in the midst of other astrophysical signals.

It is also worth mentioning that this approach is a valuable technique for high-cadence microlensing surveys including those currently underway like OGLE, MOA and KMTNet \citep{  udalski1994optical,1999PThPS.133..233M,kim2010technical}. This method will enable fast evaluation of thousands of light curves and allows astronomers to focus on high-priority light curves. It will also help shorten the time of full analysis of each light curve by providing estimations for the initial guesses of the fits.

\bibliographystyle{apj}
\bibliography{references}
 \clearpage
 
\end{document}